\documentclass[12pt]{article}
\usepackage{amsfonts}
\usepackage{graphicx}
\usepackage{amsmath}
\usepackage{float}
\newcommand {\be}{\begin{equation}}
 \newcommand {\ee}{\end{equation}}
 \newcommand {\bea}{\begin{array}}
 
 \newcommand {\eea}{\end{array}}

\textwidth=6.5in \hoffset=-.75in \voffset=-1in \numberwithin{equation}{section}
\numberwithin{figure}{section}

\begin{document}

\begin{titlepage}
\vspace{1cm}
\begin{center}
{\Large \bf {Hidden Conformal Symmetry of Kerr-Bolt Spacetimes}}\\
\end{center}
\vspace{2cm}
\begin{center}
A. M. Ghezelbash$^*${\footnote{masoud.ghezelbash@usask.ca}},  V. Kamali$^\dagger$\footnote{vkamali1362@gmail.com},
M. R. Setare$^\dagger$\footnote{rezakord@ipm.ir}
\\
$^*$Department of Physics and Engineering Physics, \\ University of Saskatchewan, \\
Saskatoon, Saskatchewan S7N 5E2, Canada\\
$^\dagger$ Department of Science of Bijar, Kurdestan University,  \\
Bijar, Iran
\vspace{1cm}
\end{center}

\begin{abstract}
We extend the recent proposal of hidden conformal symmetry to the four-dimensional spacetimes with rotational parameter and NUT twist. We consider the wave equation of a massless scalar field in background of Kerr-Bolt spacetimes and show in the ``near region", the wave equation can be reproduced by the $SL(2,R)_L \times SL(2,R)_R$ Casimir quadratic operators. 
Moreover, we compute the microscopic entropy of the dual CFT by Cardy formula and find a perfect match to Bekenstein-Hawking entropy of Kerr-Bolt spacetimes. The absorption cross section of a near region scalar field also matches to microscopic absorption cross section of the dual CFT. 
\end{abstract}
\end{titlepage}\onecolumn 
\bigskip 

\section{Introduction}
In the context of proposed Kerr/CFT correspondence \cite{stro}, (see also \cite{Solo} and \cite{Bar}) the microscopic entropy of four-dimensional extremal Kerr black hole is calculated by studying the dual chiral conformal field theory associated with the diffeomorphisms of near horizon geometry of the Kerr black hole. These diffeomorphisms preserve an appropriate boundary condition at infinity. 

The Kerr/CFT correspondence has been used in \cite{LU} and \cite{cvetic} to find the entropy of dual CFTs to 
four and higher dimensional Kerr black holes in AdS spacetimes and gauged supergravity as well as five-dimensional BMPV black holes in  \cite{ISO}. Moreover the correspondence has been used in string theory D1-D5-P and BMPV black holes in \cite{Aze} 
and in the five dimensional Kerr black hole in G\"{o}del universe \cite{G1}.
The continuous approach to the extremal Kerr black hole is essential in the proposed correspondence. For example, in the case of Reissner-Nordstrom black hole the approach to extremality is not continuous \cite{car}. The rotating bubbles, Kerr-Newman black holes in (A)dS spacetimes and rotating NS5 branes have been also considered in \cite{Aze2}, \cite{Stro2} and \cite{Naka}. The four-dimensional Kerr-Sen black hole and Kerr-Bolt spacetimes have been considered in \cite{GH2},\cite{GH3}. In all these extreme black holes, the dual chiral CFT contains a left-moving sector only. For the other works done on Kerr/CFT correspondence see \cite{KCFT}. 

Subsequently, extending Kerr/CFT correspondence to the near-extremal Kerr black hole was considered in \cite{2} (see also \cite{3}). This task is quite challenging since the right-moving sector of dual CFT turns on and no consistent boundary conditions that allow for both left and right-moving sectors, have been found yet. In this regard, an alternate approach to overcome the problem was proceeded in \cite{2}, along the way which was originally followed in \cite{MalStro} and led to AdS/CFT correspondence \cite{ads1,ads2,ads3,ads4,ads5,ads6}. In this approach, the near extreme black hole absorption cross section is entirely reproduced by two-point correlation function of a 2D non-chiral CFT.
The main progresses are made essentially on the extremal and near extremal limits in which the black hole near horizon geometries consist a certain AdS structure and the central charges of dual CFT can be obtained by analyzing
the asymptotic symmetry following the method in \cite{bh} or by calculating the boundary stress tensor of the $2D$ effective action \cite{4}.
Recently, Castro, Maloney and Strominger \cite{cms} have given evidence that the physics of Kerr black holes might be captured by a hidden conformal field theory. The authors have discussed that the existence of conformal invariance in a near horizon geometry is not a necessary condition, instead the existence of a local conformal invariance (named as hidden conformal symmetry) in the solution space of the wave equation for the propagating field is sufficient to ensure a dual CFT description. 
The analysis has been extensively applied to the wave equation of propagating scalar field in a variety of different backgrounds \cite{othe}. 

Motivated by these works on extremal and near-extremal Kerr/CFT correspondences; in this paper, we consider the four-dimensional spacetimes with rotational parameter and NUT twist and show they have a hidden conformal symmetry. 
The spacetimes with NUT twist have been studied extensively in regard to their conserved charges, maximal mass conjecture and D-bound in \cite{TN1}.

More explicitly, in section \ref{sec:sec1scalarfield}, we consider the wave equation of a massless scalar field in the background of Kerr-Bolt spacetimes and find simple angular equation for low frequency scalar field in the ``near region". In section \ref{sec:sec2hidden}, we find in the near region, the radial part of wave equation can be written in terms of $SL(2,R)_L \times SL(2,R)_R$ Casimir operators. Moreover, we find the microscopic entropy of the dual CFT and compare it to the macroscopic Bekenstein-Hawking entropy of the Kerr-Bolt spacetimes. In section \ref{sec:crosss}, we compute the absorption cross section of a near region scalar field and find perfect match to the microscopic cross section in dual CFT. 
In general, our results in this paper provide further supporting evidence for the hidden conformal symmetry of Kerr-Bolt spacetimes. 

\section{Scalar Field in Background of Kerr-Bolt Spacetimes}
\label{sec:sec1scalarfield}

The Kerr-Bolt spacetime with NUT charge $n$ and rotational parameter $a$, is given by the line element 
\begin{eqnarray}
ds^{2}&=&-\frac{\Delta(r)}{\rho^{2}}[dt+(2n\cos(\theta)-a\sin^{2}(\theta)) d\phi]^{2}
+\frac{\sin^{2}(\theta)}{\rho^{2}}[a dt-(r^{2}+n^{2}+a^{2})d\phi]^{2}
\nonumber \\
&+& \frac{\rho^{2}}{\Delta(r)}+\rho^{2} (d\theta)^{2},
\label{intro1}
\end{eqnarray}
where
\begin{eqnarray}
\rho^{2}&=&r^{2}+(n+a\cos(\theta))^{2},\\
 \label{intro2}
\Delta(r)&=& r^2-2 M r +a^{2}-n^{2}.
\label{intro3}
\end{eqnarray} 
The Kerr-Bolt spacetime (\ref{intro1}) is exact solution to Einstein equations. The inner $r_-$ and outer $r_+$ horizons of spacetime (\ref{intro1}) are the real roots of $\Delta(r)=0$. The angular velocity and the Bekenstein-Hawking entropy are $\Omega _H=\frac{a}{r_+^2+n^2+a^2}$ and $S_{BH}=\pi(r_+^2+n^2+a^2)$, respectively.

We consider the massless scalar field $\Phi$ in background (\ref{intro1}). The Klein-Gordon (KG) equation for the massless scalar field $\Phi$ 
\begin{equation}
\Box\Phi=\frac{1}{\sqrt{-g}}\partial_{\mu}(g^{\mu\nu}\partial_{\nu})\Phi=0,
\label{intro4}
\end{equation}
can be separated by taking the expansion of scalar field as 
\begin{equation}
\Phi(t,r,\theta,\phi)=\exp(-i m\phi+i\omega t) S(\theta) R(r).
\label{intro5}
\end{equation}
Actually, the equation (\ref{intro4}) reduces to 
\begin{equation}
[\partial_{r}(\Delta(r)\partial_{r})+\frac{1}{\sin\theta} \partial_{\theta}(\sin\theta\partial_{\theta})-\frac{(C(\theta)\omega+m)^{2}}{\sin^{2}\theta}+\frac{(D(r)\omega-m a)^{2}}{\Delta(r)}] S(\theta) R(r)=0,
\label{intro6}
\end{equation} 
where the functions $C(\theta)$ and $D(r)$ are given by
\begin{eqnarray}
C(\theta)&=&2n\cos(\theta)-a\sin^{2}(\theta),\\
D(r)&=&r^{2}+n^{2}+a^{2}.
\label{intro7}
\end{eqnarray}
The separated equation for $S(\theta)$ is given by
\begin{equation} 
[\frac{1}{\sin(\theta)} \partial_{\theta}(\sin(\theta)\partial_{\theta})+f(\theta)] S(\theta)=-\lambda S(\theta),\label{intro8}
\end{equation} 
where
\begin{equation}
f(\theta)=\frac{-4n^2\omega^2-m^2-4mn\omega\cos(\theta)}{\sin^{2}(\theta)}+a^2\omega^2\cos^2(\theta)+2na\omega^2\cos(\theta)~
\label{intro9}
\end{equation}
and $\lambda$ is the separation constant.
We notice from equation (\ref{intro8}) that it is possible to find a range of parameters such that the terms of order $\omega$ and $\omega^2$ can be neglected. In fact, considering low frequency scalar field and in the near region of geometry 
\begin{eqnarray}
\omega M, \omega n &\ll& 1, \\
 r &\ll& \frac{1}{\omega},
\label{intro10}
\end{eqnarray} 
the angular equation (\ref{intro8}) significantly simplifies and we get  
\begin{equation} 
[\frac{1}{\sin(\theta)} \partial_{\theta}(\sin(\theta)\partial_{\theta})-\frac{m^2}{sin^2\theta}] S(\theta)=-l(l+1)S(\theta).\label{intro88}
\end{equation} 
Equation (\ref{intro88}) is just the standard Laplacian on sphere where we choose the separation constant $\lambda=l(l+1)$ \cite{cms}.
In these limits, the radial part of KG equation becomes
\be \begin{array}{cc}
[\partial_{r}(\Delta\partial_{r})+\frac{-4Mmar\omega+m^{2}a^{2}-4n^{2}ma\omega}{\Delta(r)}+(\Delta+4(Mr+n^{2})-a+4n)
\omega^{2}]R(r)=l(l+1)R(r),
\end{array} \label{intro11}\ee
that can be rewritten as 
 \be \begin{array}{cc}
 [\partial_{r}(\Delta\partial_{r})+\frac{-4Ma\omega r_{+}+m^{2}a^{2}-4n^{2}ma\omega}{(r-r_{+})(r-r_{-})}-\frac{-4Ma\omega r_{-}+m^{2}a^{2}-4n^{2}ma\omega}{(r-r_{+})(r-r_{-})}]R(r)=l(l+1)R(r)~~
 \end{array} \label{intro13}\ee

\section{Hidden Conformal Symmetry of the Near Region Scalar Field}
\label{sec:sec2hidden}

In this section, we search for the hidden conformal symmetry of the scalar field in near region. 

We define the conformal coordinates $\omega^+, \omega^-$ and $y$ in terms of coordinates $t,r$ and $\phi$, given by
\begin{eqnarray}
 \omega^{+}&=&\sqrt{\frac{r-r_{+}}{r-r_{-}}}\exp(2\pi T_{R}\phi+2n_{R}t),
  \label{intro14}\\
  \omega^{-}&=&\sqrt{\frac{r-r_{+}}{r-r_{-}}}\exp(2\pi T_{L}\phi+2n_{L}t),
  \label{intro15}\\
  y&=&\sqrt{\frac{r_{+}-r_{-}}{r-r_{-}}}\exp(\pi( T_{R}+T_{L})\phi+(n_{R}+n_{L})t),
\label{intro16}
 \end{eqnarray} 
 where
\begin{equation}
 T_{R}=\frac{r_{+}-r_{-}}{4\pi a}, ~~~~~~T_{L}=\frac{r_{+}+r_{-}}{4\pi a}+\frac{n^{2}}{2\pi a M},
\label{intro17}
\end{equation}
and $n_{R}=0,n_{L}=-\frac{1}{4M}$.
We also define the left and right moving vector fields 
\begin{equation}
H_1=i\partial_{+},~~~~H_0=i(\omega^{+}\partial_{+}+\frac{1}{2}y
\partial_{y}),~~~~~H_{-1}=i((\omega^{+})^2\partial_{+}+\omega^{+}y\partial_{y}-y^2\partial_{-}),~~~~
\label{intro19}
\end{equation}
and
\begin{equation}
\overline{H}_1=i\partial_{-},~~~~\overline{H}_0=i(\omega^{-}\partial_{-}+\frac{1}{2}y\partial_{y}),
~~~~~\overline{H}_{-1}=i((\omega^{-})^2\partial_{-}+\omega^{-}y\partial_{y}-y^2\partial_{+}),
\label{intro20}
\end{equation}
respectively. 
The vector fields (\ref{intro19}) satisfy the $SL(2,R)$ algebra
\begin{equation}
 ~~[H_0,H_{\pm1}]=\mp i H_{\pm 1},~~~~~~~~[H_{-1},H_1]=-2iH_0,~~
\label{intro21}
\end{equation}
and similarly for $\overline{H}_1,\overline{H}_0$ and $\overline{H}_{-1}$. In terms of coordinates $t,r$ and $\phi$, the vector fields (\ref{intro19}) are
\begin{eqnarray}
 H_1&=&i\exp(-2\pi T_{R}\phi)(\Delta^{\frac{1}{2}}\partial_{r}+\frac{1}{2\pi T_{R}}\frac{r-M}{\Delta^{\frac{1}{2}}}\partial_{\phi}+\frac{aM}{2\pi T_{R}}\frac{(M+\frac{n^2}{M})r-a^2}{\Delta^{\frac{1}{2}}}\partial_{t}),~~~~~
\label{intro22}\\
H_0&=&\frac{i}{2\pi T_{R}}\partial_{\phi}-2i\frac{M^2}{aT_{R}}\partial_{t},~~~
\label{intro23}\\
H_{-1}&=&i\exp(+2\pi T_{R}\phi)(-\Delta^{\frac{1}{2}}\partial_{r}+\frac{1}{2\pi T_{R}}\frac{r-M}{\Delta^{\frac{1}{2}}}\partial_{\phi}+\frac{aM}{2\pi T_{R}}\frac{(M+\frac{n^2}{M})r-a^2}{\Delta^{\frac{1}{2}}}\partial_{t}).~~~
\label{intro24}
\end{eqnarray} 
Similarly, the vector fields (\ref{intro20}) in terms of coordinates $t,r$ and $\phi$, are given by
 \begin{eqnarray}
\overline{H}_1&=&i\exp(-2\pi T_{L}\phi-2n_{L}t)(\Delta^{\frac{1}{2}}\partial_{r} -\frac{a}{\Delta^{\frac{1}{2}}}\partial_{\phi}-2\frac{Mr+n^2}{\Delta^{\frac{1}{2}}}\partial_{t}),~~
\label{intro25}\\
\overline{H}_0&=&-2iM\partial_{t},~~~
\label{intro26}\\
\overline{H}_{-1}&=&i\exp(2\pi T_{L}2n_{L}t)(-\Delta^{\frac{1}{2}}\partial_{r} -\frac{a}{\Delta^{\frac{1}{2}}}\partial_{\phi}-2\frac{Mr+n^2}{\Delta^{\frac{1}{2}}}\partial_{t}).~~~~~
\label{intro27}
\end{eqnarray}
The quadratic Casimir operator, in terms of coordinates $\omega^+, \omega ^-$ and $y$, is given by
 \begin{equation}
H^2=-H_{0}^2+\frac{1}{2}(H_1H_{-1}+H_{-1}H_{1})=\frac{1}{4}
(y^2\partial_{y}^2-y\partial_{y})+y^2\partial_{+}\partial_{-}.
\label{intro28}
\end{equation}
In terms of coordinates $t,r,\theta,\phi$, the Casimir operator (\ref{intro28}) (as well as the other Casimir operator $\widetilde{H}^2$, constructed from (\ref{intro25}), (\ref{intro26}) and (\ref{intro27})) reduces to the radial equation (\ref{intro13}),
 \be \begin{array}{cc}
H^2R(r)=\widetilde{H}^2R(r)=l(l+1)R(r)~~~
\end{array} \label{intro29}\ee
We should note (similar to \cite{cms}) the vector fields (\ref{intro22})-(\ref{intro27}) generate only a local $SL(2,R)_L \times SL(2,R)_R$ hidden conformal symmetry for the solution space of near region KG equation in Kerr-Bolt geometry. These vectors are not globally defined since they are not periodic under $\phi \sim \phi +2\pi$ identification. The existence of local $SL(2,R)_L \times SL(2,R)_R$ hidden conformal symmetry suggests that we assume the dynamics of the near region can be described by a dual CFT. To verify this assumption, we try to find the microscopic entropy of the dual CFT. We use the Cardy formula for the entropy of a CFT
 \be \begin{array}{cc}
S_{CFT}=\frac{\pi^2}3({c_{L}T_{L}+c_{R}T_{R}}).~~~
\end{array} \label{intro30}\ee
The central charges of dual CFT were obtained in \cite{GH3}, based on analysis of the asymptotic symmetry group and given by
 \be \begin{array}{cc}
c_{R}=c_{L}=12Ma.
\end{array} \label{intro31}\ee
From the central charges (\ref{intro31}) and temperatures (\ref{intro17}), we get
 \be \begin{array}{cc}
S_{CFT}=2\pi M(r_{+}+\frac{n^2}{M})
\end{array} \label{intro32}\ee
which is in complete agreement with the Bekenstein-Hawking entropy of Kerr-Bolt spacetimes given by $\pi (r_+^2+a^2+n^2)$ \cite{entKB}, upon substitution $r_+=M+\sqrt{M^2+n^2-a^2}$. 

\section{Absorption Cross Section of Near Region Scalars}
\label{sec:crosss}

In this section, we show the absorption cross section of scalars in the near region of Kerr-Bolt spacetime can be given by dual CFT.
We rewrite the radial part of KG equation (\ref{intro11}) in terms of a new coordinate $Z$, given by
\be \begin{array}{cc}
Z=\frac{r-r_{-}}{r-r_{+}}.
\end{array} \label{intro33}\ee
In terms of coordinate $Z$, the equation (\ref{intro11}) becomes
\be \begin{array}{cc}
(1-z)z\partial_{z}R(z)+(1-z)\partial_{z}R(z)+[\frac{(2M(r_{+}+\frac{n^2}{M})\omega-ma)^2}{z(r_{+}-r_{-})^2}-
\frac{(2M(r_{-}+\frac{n^2}{M})\omega-ma)^2}{z(r_{+}-r_{-})^2}-\frac{l(l+1)}{1-z}]R(z)=0.
\end{array} \label{intro34}\ee
The solutions to (\ref{intro34}) at the outer boundary behave as
\be \begin{array}{cc}
R\simeq C_{1}r^{l}+C_{2}r^{-(l+1)}
\end{array} \label{intro35}\ee
where
\be \begin{array}{cc}
C_{1}=\frac{\Gamma(1-2i\frac{2M(r_{+}+\frac{n^2}{M})\omega-ma}{r_{+}-r_{-}})\Gamma(1+2l)}{\Gamma(1+l-2iM\omega)
\Gamma(1+l-\frac{2iM(r_{+}+r_{-}+\frac{ 2 n^2}{M})\omega-2ma}{r_{+}-r_{-}})}.
\end{array} \label{intro36}\ee
The absorption cross section is
\be \begin{array}{cc}
P_{abs}\sim \frac{1}{|C_{1}|^2} \sim\sinh(\frac{4\pi M(r_{+}+\frac{n^2}{M})\omega-2\pi ma}{r_{+}-r_{-}})|\Gamma(1+l-2iM\omega)|^2|\Gamma(1+l-\frac{2iM\omega(r_{+}+r_{-}+\frac{ 2 n^2}{M})-2ma}{r_{+}-r_{-}})
|^2.
\end{array} \label{intro37}\ee
To see explicitly that $P_{abs}$ matches to the microscopic cross section of dual CFT, we need to identify the related parameters. From the first law of thermodynamics
\be \begin{array}{cc}
T_{H}\delta S=\delta M -\Omega\delta J,
\end{array} \label{intro38}\ee
one can compute the conjugate charges as
\be \begin{array}{cc}
\delta S_{BH}=\delta S_{CFT}=\frac{\delta E_{L}}{T_{L}}+\frac{\delta E_{R}}{T_{R}},
\end{array} \label{intro39}\ee
so we can get
\be \begin{array}{cc}
\delta E_{L}=\frac{\delta M}{a}M(r_{+}+r_{-}+\frac{2n^2}{2M}),~~~~~\delta E_{R}=\frac{\delta M}{a}M(r_{+}+r_{-}+\frac{2n^2}{M})-\delta J.
\end{array} \label{intro40}\ee
Then the left and right moving frequencies are defined
\be \begin{array}{cc}
\omega_{L}=\frac{\omega}{a}M(r_{+}+r_{-}+\frac{2n^2}{M}),~~~~~~
\omega_{L}=\frac{\omega}{a}M(r_{+}+r_{-}+\frac{2n^2}{M})-m,
\end{array} \label{intro41}\ee
where
\be \begin{array}{cc}
\omega =\delta M,~~~~~~~~m=\delta J.
\end{array} \label{intro42}\ee
Using conformal weights $(h_{L},h_{R})=(l+1,l+1)$ and substituting eq (\ref{intro41}) into (\ref{intro37}), one can find
\be \begin{array}{cc}
P_{abs}\sim T_{L}^{2h_{L}-1}T_{R}^{2h_{R}-1}\sinh(\frac{\omega_{L}}{T_{L}}+\frac{\omega_{R}}{T_{R}})\mid\Gamma(h_{L}+i\frac{\omega_{L}}{2\pi T_{L}})\mid^2 \mid\Gamma(h_{R}+i\frac{\omega_{R}}{2\pi T_{R}})\mid^2,
\end{array} \label{intro43}\ee
which is the finite temperature absorption cross section for a $2D$ CFT.
\section{Concluding Remarks}
\label{sec:con}
We have shown in this paper that four-dimensional rotating spacetimes with NUT twist have hidden conformal symmetry in solution space of massless scalar field. Specifically, we consider the wave equation of a massless scalar field in the background of Kerr-Bolt spacetimes and find in the ``near region", the wave equation can be written in terms of $SL(2,R)_L \times SL(2,R)_R$ Casimir operators. The macroscopic entropy and the absorption cross section of near region scalar field match precisely to those of microscopic dual CFT. 
All our results in this paper provide further evidence for the hidden conformal symmetry of Kerr-Bolt spacetimes. 
There is however the issue of removing the singularities of Kerr-Bolt spacetimes. As it is known, these singularities should be removed from Kerr-Bolt spacetimes in a consistent manner \cite{GHKB}. An interesting question is how the removal of singularities can be realized in CFT side. Another interesting question is to study the wave equation of scalar field in the background of Kerr-Bolt-(A)dS spacetimes to find the hidden conformal symmetry. We leave these questions for future articles. \newline 

\bigskip

{\Large Acknowledgments}\newline

This work was supported in part by the Natural Sciences and Engineering Research
Council of Canada.


\begin{thebibliography}{99}
\bibitem{stro} 
M. Guica, T. Hartman, W. Song and A. Strominger, [arXiv:0809.4266].
\bibitem{Solo}
S. N. Solodukhin, \textit{Phys. Lett.} \textbf{B454} (1999) 213.
\bibitem{Bar}
G. Barnich and F. Brandt,
\textit{Nucl. Phys.} \textbf{B633} (2002) 3.
\bibitem{LU} 
H. L\"{u}, J. Mei and C.N. Pope, [arXiv:0811.2225].
\bibitem{cvetic}
D.D.K. Chow, M. Cveti\u{c}, H. L\"{u} and C.N. Pope, [arXiv:0812.2918].
\bibitem{ISO}  H. Isono, T.-S. Tai and W.-Y. Wen, [arXiv:0812.4440].
\bibitem{Aze} 
T. Azeyanagi, N. Ogawa and S. Terashima, [arXiv:0812.4883].
\bibitem{G1} 
J.-J. Peng and S.-Q. Wu, [arXiv:0901.0311].
\bibitem{car} 
S.M. Carroll, M.C. Johnson and L. Randall, [arXiv:0901.0931].
\bibitem{Aze2} 
T. Azeyanagi, N. Ogawa and S. Terashima, [arXiv:0811.4177].
%
\bibitem{Stro2}
T. Hartman, K. Murata, T. Nishioka and A. Strominger, [arXiv:0811.4393].
\bibitem{Naka}
Y. Nakayama, [arXiv:0812.2234].
\bibitem{GH2}
A.M. Ghezelbash, \textit{JHEP} \textbf {0908} (2009) 045.  
\bibitem{GH3}
A.M. Ghezelbash, [arXiv:0902.4662] 
\bibitem{KCFT}
Y. Matsuo, T. Tsukioka and C.-M. Yoo, [arXiv:1007.3634]; 
B. Carneiro da Cunha and A. R. de Queiroz, [arXiv:1006.0510];  
H. Wang, D. Chen, B. Mu and H. Wu, [arXiv:1006.0439];
B. Chen and J. Long, [arXiv:1006.0157]; 
J. Rasmussen, [arXiv:1005.2255]; 
D. Chen, P. Wang and H. Wu, [arXiv:1005.1404]; 
R. Li, M.-F. Li and J.-R. Ren, \textit{Phys. Lett.} \textbf {B691} (2010) 249; 
B. Chen, J. Long, \textit{JHEP} \textbf {1006} (2010) 018;
J. Rasmussen, [arXiv:1004.4773]; C. Krishnan, \textit{JHEP} \textbf {1007} (2010) 039; M. Becker, S. Cremonini and W. Schulgin, [arXiv:1004.1174]; 
J.-J. Peng and S.-Q. Wu, \textit{Nucl. Phys.} \textbf {B828} (2010) 273; D. Grumiller and  A.-M. Piso, [arXiv:0909.2041];
T. Hartman, W. Song and A. Strominger, \textit{JHEP} \textbf {1003} (2010) 118; J. Rasmussen, \textit{Int. J. Mod. Phys.} \textbf {A25} (2010) 1597; Y. Matsuo, T. Tsukioka and C.-M Yoo, \textit{Europhys. Lett.} \textbf {89} (2010) 60001; I. Bredberg, T. Hartman, W. Song and A. Strominger, \textit{JHEP} \textbf {1004} (2010) 019; Y. Matsuo, T. Tsukioka and C.-M. Yoo, \textit{Nucl. Phys.} \textbf {B825} (2010) 231; O.J.C. Dias, H. S. Reall and J.E. Santos, \textit{JHEP} \textbf {0908} (2009) 101; L.-M. Cao, Y. Matsuo, T. Tsukioka and C.-M. Yoo, \textit{Phys. Lett.} \textbf {B679} (2009) 390; W.-Y. Wen, [arXiv:0903.4030]; C. Krishnan and S. Kuperstein, \textit{Phys. Lett.} \textbf {B677} (2009) 326; 
C.-M. Chen, Y.-M. Huang, J.-R. Sun, M.-F. Wu and S.-J. Zou, [arXiv:1006.4097]; 
C.-M. Chen and J.-R. Sun,[arXiv:1004.3963]; J. Mei, \textit{JHEP} \textbf {1004} (2010) 005; C.-M. Chen, Y.-M. Huang and S.-J. Zou, \textit{JHEP} \textbf {1003} (2010) 123; J.-J. Peng and S.-Q. Wu, \textit{Nucl. Phys.} \textbf {B828} (2010) 273; C.-M. Chen, J.-R. Sun and S.-J. Zou\textit{JHEP} \textbf {1001} (2010) 057; E. Barnes, D. Vaman and C. Wu, \textit{Class. Quant. Grav.} \textbf {27} (2010) 095019; X.-N. Wu and Y. Tian, \textit{Phys. Rev.} \textbf {D80} (2009) 024014; T. Azeyanagi, G. Compere, N. Ogawa, Y. Tachikawa, and S. Terashima, \textit{Prog. Theor. Phys.} \textbf {019122} (2009) 355; M. R. Garousi and A. Ghodsi, \textit{Phys. Lett.} \textbf {B687} (2010) 79; D. Astefanesei and  Y.K. Srivastava, \textit{Nucl. Phys.} \textbf {B822} (2009) 283; K. Hotta, \textit{Phys. Rev.} \textbf {D79} (2009) 104018; G. Compere, K. Murata and T. Nishioka, \textit{JHEP} \textbf {0905} (2009) 077.
\bibitem{2} I. Bredberg, T. Hartman, W. Song and A. Strominger,
[Arxive:0907.3477].
\bibitem{3} O. J. C. Dias, H. S. Reall and E. Stantos, JHEp, 0908,101, (2009); Y. Matsuo, T. Tsukioka and C.M. Yoo,   \textit{Nucl. Phys} \textbf{B825}(2010) 231; A. J. Amsel, D. Marolf and M.M. Roberts. \textit{JHEP} \textbf{0910} (2009) 021; A. Castro and F. Larsen, \textit{JHEP} \textbf{0909} (2009) 088.
\bibitem{MalStro}
J. M. Maldacena and A. Strominger, \textit{Phys. Rev.} \textbf {D56} (1997) 4975; J. M. Maldacena, J. Michelson and A. Strominger, \textit{JHEP} \textbf {9902} (1999) 011.
\bibitem{ads1}
E. Witten, \textit{Adv. Theor. Math. Phys.} \textbf {2} (1998) 253.
\bibitem{ads2}
J. Maldacena, \textit{Adv. Theor. Math. Phys.} \textbf {2} (1998) 231.
\bibitem{ads3}
K. Sfetsos and K. Skenderis, \textit{Nucl. Phys.} \textbf{B517} (1998) 179.
\bibitem{ads4}
D.Z. Freedman, S.D. Mathur, A. Matusis and L. Rastelli, \textit{Nucl. Phys.} \textbf{B546} (1999) 96.
\bibitem{ads5}
S.S. Gubser, I.R. Klebanov and A.M. Polyakov, \textit{Phys. Lett.} \textbf {B428} (1998) 105.
\bibitem{ads6}
A.M. Ghezelbash, K. Kaviani, S. Parvizi and A.H. Fatollahi, 
\textit{Phys. Lett.} \textbf{B435} (1998) 291; A.M. Ghezelbash, M. Khorrami, A. Aghamohammadi, \textit{Int. J. Mod. Phys.} \textbf{A14} (1999) 2581. 
\bibitem{bh}J.D. Brown, and M. Henneaux, Commun. Math. Phys.104,207,(1986).
\bibitem{4}T. Hartman and A. Strominger. \textit{JHEP} \textbf{0904} (2009) 026.
\bibitem{cms}
A. Castro, A. Maloney and A. Strominger, \textit{Phys. Rev.} \textbf {D82} (2010) 024008.
\bibitem{othe}
M.R. Setare, V. Kamali, [arXiv:1008.1123]; 
K.-N. Shao and Z. Zhang, [arXiv:1008.0585]; 
B. Chen, J. Long and J.-j. Zhang, [arXiv:1007.4269]; 
Y. Matsuo, T. Tsukioka and C.-M. Yoo, 
[arXiv:1007.3634];
R. Li, M.-F. Li and J.-R. Ren, [arXiv:1007.1357]; 
A.P. Porfyriadis and F. Wilczek, [arXiv:1007.1031]; 
R. Monteiro, [arXiv:1006.5358]; 
C.-M. Chen, Y.-M. Huang, J.-R. Sun, M.-F. Wu and S.-J. Zou, [arXiv:1006.4097]; 
[arXiv:1006.4092]; 
H. Wang, D. Chen, H. Wu and H. Yang, [arXiv:1006.0439]; B. Chen, J. Long, [arXiv:1006.0157]; M. Becker, S. Cremonini and W. Schulgin, [arXiv:1005.3571]; 
J. Rasmussen, [arXiv:1005.2255]; C. Krishnan, [arXiv:1005.1629]; 
D. Chen, P. Wang, H. Wu and H. Yang, [arXiv:1005.1404]; 
R. Li, M.-F. Li and J.-R. Ren, \textit{Phys. Lett.} \textbf{B691} (2010) 249.
 
\bibitem{TN1}
R. Clarkson, A.M. Ghezelbash and R.B. Mann, \textit{Phys. Rev. Lett.} \textbf{91} (2003) 061301; \textit{Nucl. Phys.} \textbf{B674} (2003) 329; \textit {Int. J. Mod. Phys}
\textbf{A19} (2004) 3987. 
\bibitem{entKB}
R. Mann, \textit{Phys. Rev.} \textbf{D61} (2000) 084013.
\bibitem{GHKB}
A.M. Ghezelbash, R.B. Mann and R.D. Sorkin, \textit{Nucl. Phys.} \textbf{B775} (2007) 95.

\end{thebibliography}
\end{document}